\documentclass[12pt]{article}

\newcommand{\bx}{{\bf x}}
\newcommand{\bk}{{\bf k}}
\newcommand{\bp}{{\bf p}}
\newcommand{\bnab}{{\bf \nabla}}
\newcommand{\fith}{\widehat\varphi}
\newcommand{\fiti}{\widetilde\phi}

\begin{document}
\mbox{} \hfill UCLA/00/TEP/27
\begin{center}{\Large \bf Alternative Dimensional Reduction via the Density Matrix}\\
[.3in]
 C. A. A. de Carvalho$^{a,b}$\footnote{aragao@physics.ucla.edu, aragao@if.ufrj.br}
 and J. M. Cornwall$^a$\footnote{cornwall@physics.ucla.edu}\\
{\it $^a$Department of Physics and Astronomy, 
University of California, Los Angeles, CA 90095-1547\\
$^b$ Permanent address: Instituto de F\'\i sica,
UFRJ, 
C.P.~68528, Rio de Janeiro, RJ 21945-970, Brasil}\\

A. J. da Silva\footnote{ajsilva@if.usp.br}\\
{\it Instituto de F\'\i sica, Universidade de S\~ao Paulo, \\ 
C.P.~66318, S\~ao Paulo, SP 05315-970, Brasil}
\date{\today}
\end{center}

\begin{abstract}

We give graphical rules, based on earlier work for the functional Schr\"odinger equation, for constructing the density matrix for scalar and gauge fields in equilibrium at finite temperature $T$.  More useful is a dimensionally-reduced effective action (DREA) constructed from the density matrix by further functional integration over the arguments of the density matrix coupled to a source.  The DREA is an effective action in one less dimension which may be computed order by order in perturbation theory or by dressed-loop expansions; it encodes all thermal matrix elements.  We term the DREA procedure {\em alternative dimensional reduction}, to distinguish it from the conventional dimensionally-reduced field theory (DRFT) which applies at infinite $T$.   The DREA is useful because it gives a dimensionally-reduced theory useable at any $T$ including infinity, where it yields the DRFT, and because it does not and cannot have certain spurious infinities which sometimes occur in the density matrix itself or the conventional DRFT; these come from $\ln T$ factors at infinite temperature.  The DREA can be constructed to all orders (in principle) and the only regularizations needed are those which control the ultraviolet behavior of the zero-$T$ theory.  An example of spurious divergences in the DRFT occurs in d=2+1 $\varphi^4$-theory dimensionally reduced to d=2.  We study this theory and show that the rules for the DREA replace these ``wrong" divergences in physical parameters by calculable powers of $\ln T$; we also compute the phase transition temperature of this $\varphi^4$-theory in one-loop order. Our density-matrix construction is equivalent to a construction  of the Landau-Ginzburg `` coarse-grained free energy'' from a microscopic Hamiltonian.  
    
\end{abstract}


\section{Introduction}
\label{introduction}

The study of relativistic field theories at finite $T$ \cite{finitet} and/or density \cite{books}, in or out of equilibrium \cite{noneq}, is quite mature but continues to be of great interest, for example, in connection with RHIC and CERN relativistic heavy-ion collision experiments searching for the QCD transition from a hadronic to a quark-gluon plasma phase; a possible color-flavor locking transition to a superconducting phase of QCD at finite particle density; the transition to a disoriented chiral condensate of pions that could provide an explanation for the so-called Centauro events observed in cosmic rays; and cosmological transitions \cite{phenom}.  There is an equally-extensive body of literature on non-relativistic systems at finite $T$ and so forth, the preponderance of which deals with critical phenomena \cite{critph}. 

In this paper we will be concerned with the description of all these systems via the density matrix and a related dimensionally-reduced effective action; the methods used can be easily extended to the closely-related Wigner distribution function.\footnote{The density matrix is a diagonal matrix element; the Wigner distribution function is effectively an off-diagonal matrix element.  The reader will see that it is quite straightforward to extend the density-matrix rules to the Wigner distribution function; we will not discuss the Wigner distribution function rules in detail here.}  We begin by finding rules for describing the density matrix as the negative exponent of a dimensionally-reduced action $S_D$. These rules follow from a microscopic theory by the usual procedure of integrating over Euclidean time evolution.  This action, however, is not our final goal, because it may contain spurious ultraviolet divergences. If spurious divergences do occur, they will also occur in the usual dimensionally-reduced field theory (DRFT) arrived at by the standard procedure \cite{dimred} of going to infinite $T$.  So while it is interesting to know how to construct the density matrix systematically, it is of less use than the ultimate result of our procedure, which is a dimensionally-reduced effective action (DREA) constructed by further functional integration over the density matrix, coupled to sources.  The DREA has no spurious divergences and contains in it all possible thermal matrix elements which could be constructed by conventional means.   The procedure adopted here is a natural extension of a construction used \cite{cornwall} in the functional Schr\"odinger equation. 

In the case of d=0+1 theories (quantum mechanics) in many situations the integral over the Euclidean time evolution can be done explicitly \cite{semi,bsas,pre}, yielding integrals which are complicated functions of $T$;  in field theory it is, of course, impossible to do the corresponding Euclidean time integrals exactly.  But one must expect similarly-complicated dependences on $T$ in the field-theoretic effective action, and these dependences will be important in estimating such properties of thermal field theories as phase transition temperatures.

For relativistic systems it is common to use the DRFT, which is appropriate in principle only for infinite $T$, at temperatures very high compared to typical energies of the zero-$T$ field theory.  The DRFT may (we will study an example) have divergences which are not present at any finite $T$; while this does not impair the essential correctness of the dimensionally-reduced theory it does introduce arbitrary parameters which should not be there.  The point is that at any $T$, however large, a field theory may have no ultraviolet divergences not already contained in the zero-$T$ field theory; this well-known consideration follows from the Feynman rules for finite-$T$ field theory.  But sometimes a dimensionally-reduced field theory does have ``wrong" divergences (which must, in fact, be absent in physical quantities)  which interfere with the use of the DRFT as a good approximation for large but finite $T$.  An example is finite-$T$ $d$=2+1 $\varphi^4$ theory, where the dimensionally-reduced theory is just $d$=2 $\varphi^4$ theory.  This latter theory has a one-loop logarithmic mass divergence which is absent in  the $d$=2+1 theory, and we will give rules for the density matrix for the $d$=2+1 theory which automatically remove this divergence (replacing it by a finite and calculable factor of $\ln T$).  

For nonrelativistic systems there are corresponding problems of renormalization, but they are generally bypassed in the study of second-order phase transitions.    In studying such transitions, the need for ultraviolet renormalization and the fact that one has to start from a microscopic theory can be relegated to secondary status, because one is concerned only with long-range properties. Most treatments of nonrelativistic systems exhibiting phase transitions, such as in condensed matter physics, are not concerned with a detailed microscopic knowledge of the Hamiltonian, but rather with a universal description via the renormalization group that captures the essence of the long-range collective behavior responsible for the transition and so renormalization is not intrinsically required. 
The use of the renormalization group in such systems is a tool to extract the long-range behavior: it provides a way of eliminating short-range scales in favor of collective long-range parameters by integrating out short-distance scales to obtain effective Hamiltonians, and investigating the appearance of fixed points associated with criticality. Near those fixed points, it allows for the calculation of the exponents that characterize the critical behavior of physical quantities.  The theories studied can be (and often should be) defined on a lattice; whether this is the correct microscopic description or not makes no difference to the long-range properties.  In other words, the tools used to study long-range behavior are not capable of calculating parameters such as phase-transition temperatures, whose values depend on phenomena at shorter distances and in particular on the detailed functional dependence of the Hamiltonian on $T$.  This Hamiltonian is typically 
 a Landau-Ginzburg phenomenological ``coarse-grained free energy'', whose parameters' dependence on $T$ and, possibly, on external parameters is chosen purely phenomenologically; it is not derived from a fundamental picture.

This article addresses the question of how actually to derive the DREA, starting from the microscopic Hamiltonian that describes a field theory. This question is essentially equivalent to asking how to find the Landau-Ginzburg coarse-grained free energy, as the dimensionally-reduced theory whose matrix elements are those of the DREA itself.  One can thereby replace the phenomenology of the Landau-Ginzburg free energy by explicit calculation of, say, the $T$-dependence of this free energy.  In the case we study in this paper the underlying microscopic theory is considered to be known (a relativistic quantum field theory), and the density matrix and the DREA are to be found.  We will consider scalar and gauge field theories; nothing much new is found by studying fermions\footnote{Except possibly for topological effects involving fermions, which we  will not discuss here.}.
Just as with any effective action, the DREA contains in its infinitely-many terms all the (one-particle irreducible) matrix elements which can be constructed by tracing time-independent operator products with the density matrix.
But actually constructing the density matrix itself is not as useful, since this object contains the same spurious infinities that the DFRT contains.

The technical reason why these spurious infinities occur is that the propagators to be used in integrating over Euclidean time dynamics to form the density matrix are not the standard thermal propagators, and these non-standard propagators do lead to spurious ultraviolet divergences. However, in constructing the partition function from the density matrix, the density-matrix propagators combine with integration over certain source terms involving the time-independent fields which are the argument of the density matrix and cancellations occur; the result is equivalent to using the usual thermal propagators.  As a result, the partition function can only involve ultraviolet divergences (and subsequent renormalization) of the $T=0$ $d+1$-dimensional theory.  The same combination of propagators and source terms occurs in constructing the DREA, which therefore has only the $T=0$ divergences.  
 
We note here some general properties of the DREA:  At $T$ large compared to any mass scale $m$ in the original zero-temperature theory, {\em local} terms\footnote{Local terms in an effective action are candidates for the action of a field theory which could have the DREA as its effective action.  Non-local terms correspond to one-particle irreducible Green's functions.} in the DREA with large powers of its argument $\phi(\bx )$ will be accompanied by large negative powers of $T$ and vanish at infinite temperature; this is the way that local terms in the DREA yield the DRFT.  As $T\to 0$, there will also be factors like $\exp (-m/T)$ along with these powers of $T$, and the zero-temperature limit is well-defined.  In this limit the DREA is precisely the (negative logarithm of) the ground-state Schr\"odinger wave functional \cite{cornwall}.  The point which is of particular interest in the present paper is the appearance of marginal terms depending on $\ln T$ which lead to the unwanted divergences in the DRFT at infinite temperature.

The paper is organized as follows: Section \ref{densitymatrix} defines the density matrix for a scalar field theory, shows how to compute the reduced theory perturbatively, discusses how its ultraviolet behavior is related to the renormalization of the original theory, and outlines the construction of the DREA.  Section \ref{gauge} discusses some general questions of gauge invariance of the DREA for gauge theories. Section \ref{applications} illustrates how the dimensional reduction obtained via density matrices can be used in practical applications: a modified reduced theory (defined by its Feynman rules) is proposed, which should yield the same results as the one derived from the density matrix, and a perturbative discussion of the phase transition for a scalar theory in two spatial dimensions is presented. This last application is intended to emphasize that our approach reduces the discussion of the phase transition to the study of an (almost) ordinary system in $d$ dimensions, plus the $d+1$ subtractions required by the usual renormalization procedure.      
                  

\section{The density matrix for scalar theories}    
\label{densitymatrix}

The partition function for a self-interacting scalar field theory
in contact with a thermal reservoir at $T$ ($\beta=1/T$)
can be written as a functional integral over the density matrix $\rho$:
\begin{equation}
\label{Z1}
Z(\beta)=\int [{\cal D}\phi]\,\,\rho[\beta;\phi,\phi]\;,
\end{equation}
\begin{equation}
\label{varphi}
\rho[\beta;\phi,\phi]=\oint
[{\cal D}\varphi]\, e^{-S[\varphi]}\;,
\end{equation}
\begin{equation}
S[\varphi]=\int_0^{\beta} d\tau \int d^d\bx\, ({\cal L}_F[\varphi]+{\cal L}_I[\varphi])\;,
\end{equation}
\begin{equation}
{\cal L}_F[\varphi]=\frac{1}{2} \left\{(\partial_{\tau} \varphi)^2 + (\bnab \varphi)^2 + m^2 \varphi^2 \right\}\;.
\end{equation}
Here ${\cal L}_F$ and ${\cal L}_I$ are, respectively, the free and interacting Lagrangeans. We will be interested in $\varphi^4$ interactions (both broken and unbroken). The field $\phi(\bx)$ is the boundary value of $\varphi(\tau,\bx)$ at both $\tau=0$ and $\tau=\beta$, that is, the integral $\oint$ is to be performed over all $\varphi$ that satisfy the boundary conditions $\varphi(0,\bx)=\varphi(\beta,\bx)=\phi(\bx)$. The density matrix is a functional of $\phi$ only. The remaining integral over the $\phi$'s is unrestricted (except for the vacuum boundary conditions that must be imposed at spatial infinity). 

We may write $\rho = e^{-S_D}$, $S_D$ being a certain $T$-dependent dimensionally-reduced action.  This action is NOT the DREA, whose construction needs further discussion.  The field $\phi$ which is the argument of $S_D$ depends only on the $d$ spatial coordinates; all the $\tau$ dependence of the original $d+1$-theory has been eliminated through the $\varphi$ integration. The fields $\phi(\bx)$ are the natural degrees of freedom of the reduced theory.  Any thermal observable can be constructed by integrating over the fields $\phi(\bx )$ an appropriate function of $\phi$ weighted with the corresponding diagonal element of the density matrix. The Euclidean time evolution can be viewed as an intermediate step which calculates the weights.  In carrying out the functional integrals over $\phi$, one would notice certain cancellations between divergences occurring in the construction of $S_D$ and divergences in the functional integral.  These cancellations remove the spurious ultraviolet divergences (those not occurring in the $T=0$ theory).  

To complete the construction of the DREA it is only necessary to introduce further sources $J(\bx )$ coupled to the field $\phi(\bx )$ through the usual term $\int J\phi$.  The logarithm of the resulting functional integral over $\phi$ is then Legendre-transformed to yield the DREA.

Note that neither the density matrix nor the DREA is, in general, of the form $e^{-\beta H}$ with $H$ being independent of $\beta$ as in ordinary statistical mechanics; their $\beta$ dependence is far more complicated (which justifies the ``almost ordinary'' of the previous Section). This had already been pointed out in the analogous discussion of the transfer matrix carried out in Ref. \cite{wilsonkogut}. The density matrix provides a direct but alternative way of deriving a dimensionally-reduced theory. We will proceed to construct it perturbatively, and indicate where appropriate the generally straightforward generalization to dressed-loop expansions.


\subsection{The perturbative construction for scalars}
\label{pertconst}

The integral over the fields $\varphi$ in equation (\ref{varphi}) has to respect the boundary conditions. For the free theory, the integral is quadratic and can be computed exactly by the saddle-point method. We must solve the free field equation of motion, subject to the boundary conditions:
\begin{equation}
\label{EMBC}
\label{EM}
(-\partial_{\tau}^2 - \bnab^2 + m^2)\varphi(\tau,\bx)=0\;,
\end{equation}
\begin{equation}
\label{BC}
\varphi(0,\bx)=\varphi(\beta,\bx)=\phi(\bx)\;.
\end{equation}
Fourier transforming in $\bx$ leads to an ordinary differential equation. The solution satisfying the boundary conditions is:
\begin{equation}
\label{PHITILDE}
\fith(\tau,\bx)=\int \frac{d^d \bk}{(2\pi)^d}\, \frac{\cosh[w_{\bk}(\tau - \beta/2)]}{\cosh(\beta w_{\bk}/2)}\, \fiti(\bk)\,
 e^{i\bk\cdot\bx}\;
\end{equation}
\begin{equation}
\label{PHIK}
\phi(\bx)\equiv \int \frac{d^d \bk}{(2\pi)^d}\, \fiti(\bk)\,
 e^{i\bk\cdot\bx}\;,
\end{equation}
where $w_{\bk}=+[\bk^2+m^2]^{1/2}$. It depends functionally on $\phi(\bx)$. Its $\tau$ dependence, however, is completely specified.

We now expand $\varphi$ around $\fith$, treating it as a thermal background for the free theory:
\begin{equation}
\label{PHI+ETA}
\varphi(\tau,\bx)=\fith(\tau,\bx) + \eta(\tau,\bx)\;,
\end{equation}
\begin{equation}
\label{BCETA}
\eta(0,\bx)=\eta(\beta,\bx)=0\;.
\end{equation}
(Similarly, the $\varphi$ propagator is the sum of the $\fith$ propagator and the $\eta$ propagator.)
As usual, the fluctuation $\eta$ has to vanish at $\tau=0$ and $\tau=\beta$ because $\fith$ already satifies the boundary conditions. The expression for the density matrix element becomes:
\begin{equation}
\label{RHO}
\rho[\beta;\phi,\phi]\equiv e^{-S_D}= e^{-S_F[\fith]}\oint [{\cal D}\eta] e^{-S_F[\eta]-S_I[\fith+\eta]}\;,
\end{equation}
where $S_F$ and $S_I$ refer to free and interacting actions, respectively, and the path integral sums over fluctuations that vanish at $\tau=0$ and $\tau=\beta$.  

Note that equation (\ref{RHO}) is exactly of the same functional form as given for the functional Schr\"odinger equation in Ref. \cite{cornwall}; it is a $d+1$-dimensional functional integral (with an extra factor $\exp -S_F[\fith]$), with sources depending on $\fith$ through the interaction term $S_I[\fith+\eta]$.  The free propagators used to evaluate the functional integral differ from those of the functional Schr\"odinger equation, but the structure is exactly the same, and so the remarks made in \cite{cornwall} concerning the extension of perturbation theory to dressed-loop expansions apply without change to the present case.  We therefore need not repeat those remarks here.     

For the free theory, $S_I=0$ and we can compute the integral exactly to obtain the fluctuation determinant (to power $-1/2$) using the techniques of Refs.\ \cite{semi,bsas,pre}. For the interacting theory, we first perform a functional Taylor expansion of the interacting action around the thermal background, so as to obtain a series (or a polynomial in the $\varphi^4$ case) in powers of $\eta$. Then, we expand the exponential $e^{-S_I}$ in a power series. The integral over fluctuations will Wick-contract the various products of $\eta$'s which appear multiplied by derivatives of the interacting Lagrangean taken at the thermal background. As a consequence, one is led to compute:
\begin{equation}
\label{ETAETA}
\langle\eta(\tau_1,\bx_1)\cdots\eta(\tau_k,\bx_k)\rangle\equiv
\Delta_F^{-1/2}\oint [{\cal D}\eta] 
\,e^{-S_F[\eta]}\,\eta(\tau_1,\bx_1)\cdots\eta(\tau_k,\bx_k)\;.
\end{equation}
$\Delta_F$ stands for the fluctuation determinant of the free theory, a (infinite) normalization constant that depends on $\beta$. The result is simply:
\begin{equation}
\label{SUMP}
\langle\eta(\tau_1,\bx_1)\cdots\eta(\tau_k,\bx_k)\rangle =
\sum_P{G_\eta}(\tau_{i_1},\bx_{i_1};\tau_{i_2},\bx_{i_2})\cdots 
{G_\eta}(\tau_{i_{k-1}},\bx_{i_{k-1}};\tau_{i_k},\bx_{i_k})\;,
\end{equation}
if $k$ is even, and zero otherwise. $\sum_P$ denotes sum over all 
possible pairings of the $\{i_j\}$. There remains to compute the propagator $G_\eta$, which must satisfy:
\begin{equation}
\label{GETAEQ}
(-\partial_\tau^2 - \bnab^2 + m^2)G_\eta(\tau,\bx;\tau',\bx')
=\delta(\tau-\tau')\delta^d(\bx-\bx')\;,
\end{equation}
\begin{equation}
\label{GETABC}
G_\eta(0,\bx;\tau',\bx')= G_\eta(\beta,\bx;\tau',\bx')=0\;,
\end{equation} 
because of the vanishing boundary conditions for the fluctuations. If we Fourier transform the spatial coordinates, we obtain an ordinary differential equation which leads to \cite{cornwall,semi,bsas,pre}:
\begin{equation}
\label{GETA}
G_\eta(\tau,\bx;\tau',\bx')=\int \frac{d^d \bk}{(2\pi)^d}\, \frac{\sinh(w_{\bk}\tau_<)\sinh[w_{\bk}
(\beta-\tau_>)]}{w_{\bk}\sinh(\beta w_{\bk})}\, 
 e^{i\bk\cdot(\bx-\bx')}\;,
\end{equation}
where $\tau_<(\tau_>)\equiv{\rm min(max)}\{\tau,\tau'\}$. Note that this propagator is {\sl not} $\tau$-translation invariant. Formally, we have:
\begin{equation}
\label{RHOA}
\rho[\beta;\phi,\phi]=\rho_0[\beta;\phi,\phi]+\sum_{m=1}^\infty\rho_m[\beta;\phi,\phi]\;,
\end{equation}
\begin{equation}
\label{RHO0}
\rho_0=\Delta_F^{-1/2}e^{-S_F[\widehat\varphi]}\;,
\end{equation}
\begin{equation}
\label{RHOM}
\rho_m=\rho_0
\left[\frac{(-1)^m}{m!}\sum_{n_1,\ldots ,n_m=0}^\infty
(\prod_{j=1}^{m}\int_j 
\frac{{\cal L}_I^{(n_j)}[\widehat\varphi]}{n_j!})
\,\langle \eta^{n_1}(1)\cdots \eta^{n_m}(m)\rangle\,\right]\;,
\end{equation}
where $\int_j\equiv\int_0^{\beta} d\tau_j\int d^d\bx_j$, the argument of $\eta(j)$ stands for $(\tau_j,\bx_j)$, and ${\cal L}_I^{(n_j)}[\widehat\varphi]$ is the $n_j$-th derivative of ${\cal L}_I$, computed at the thermal background $\widehat\varphi$. Expression (\ref{RHOA}) depends functionally on $\phi$ through $\widehat\varphi$.

The spatial integrations should be converted to momentum space, where the propagator is diagonal ($\propto \delta^d(\bk'-\bk)$). Performing the integral over $\bk$ will lead to ultraviolet divergences. Therefore, we introduce a momentum cutoff $\Lambda$. It will be convenient to think of such a cutoff as analogous to an inverse lattice spacing in a lattice regularized version of the theory. In the cutoff theory, we can compute each term in the series and obtain the $T$ and cutoff dependence of the various coefficients of a functional expansion in $\phi(\bx )$ (or $\fiti(\bk)$). The logarithm of (\ref{RHOA}) will then be $-S_D$, a cutoff action which will enter further functional integrals over $\phi(\bx )$.  


\subsection{The ultraviolet behavior}
\label{uvbehavior}

The construction of the previous section allows us to find a cutoff reduced action $S_D$ order by order in perturbation theory. We will now specialize our discussion to ${\cal L}_I=\lambda\varphi^4/4!$. As  $\Lambda\to\infty$, the reduced theory exhibits the ultraviolet divergences of both $d+1$ and $d$ dimensions (the Appendix  shows this explicitly for $d=2$). When computing correlation functions, the former are to be eliminated through the usual renormalization procedure, which requires the same subtractions as in the $T=0$, $d+1$-dimensional, original theory. The $d$-dimensional divergences, however, are a consequence of failing to complete at the density-matrix stage the two-step process of integrating over the fields $\eta(\tau ,\bx   )$ followed by integration over the $\phi(\bx )$ fields which are the arguments of the density matrix. The $\eta$-generated divergences should automatically cancel upon doing the final integration over the remaining $\phi(\bx)$ which is necessary to find the partition function  $Z(\beta)$.  Indeed, this is what happens, as we will show.   There is no reason that these divergences should be absent from $S_D$ itself.

We can find these cancellations by calculating $Z(\beta)$. After having done the integration over the fluctuation fields $\eta$, we still have to integrate
(in momentum space) over the remaining $\fiti({\bp})$ in order to compute the vacuum bubbles in the expansion (\ref{RHOA}). The free part of the reduced action $S_D$ for $\fiti({\bp})$ is given by $S_F[\fith]$. An integration by parts, and use of the equation of motion satisfied by $\fith$ lead to:
\begin{equation}
\label{SFI}
S_F[\phi]=\frac{1}{2}\int \frac{d^d \bp}{(2\pi)^d}\, \fiti(\bp)[2w_{\bp}\tanh(\beta w_{\bp}/2)]\fiti(-\bp)\;,
\end{equation}
which defines the free $\fiti$-propagator as:
\begin{equation}
\label{GFI}
\widetilde{G}_\phi({\bp},{\bp}')=(2\pi)^d \delta^d ({\bp}+{\bp}')[2w_{\bp}\tanh(\beta w_{\bp}/2)]^{-1}\;.
\end{equation}
This propagator will be used to contract any two $\fiti$ fields. In the language of Feynman graphs, any line representing a contraction of the corresponding $\fith$ fields appearing in (\ref{RHO}) will have a factor:
\begin{equation}
\label{FACFI}
\frac{\cosh[w_{\bk}(\tau - \beta/2)]}{\cosh(\beta w_{\bk}/2)}\,[2w_{\bk}\tanh(\beta w_{\bk}/2)]^{-1}
\,\frac{\cosh[w_{\bk}(\tau' - \beta/2)]}{\cosh(\beta w_{\bk}/2)}\;.
\end{equation}
For contractions of $\eta$ fields, we must use the $\widetilde{G}_\eta$ propagator. Any line representing such a contraction will have a factor:
\begin{equation}
\label{FACETA}
\frac{\sinh(w_{\bk}\tau')\sinh[w_{\bk}(\tau - \beta/2)]\Theta(\tau-\tau')}{w_{\bk}\sinh(\beta w_{\bk})}
+ \;[\tau'\leftrightarrow \tau]\;.
\end{equation}
Note that this not the usual thermal propagator.
However, the combination of the integration over $\eta$ and $\fiti$ together leads to a cancellation of the ultraviolet divergences appearing in the separate integrals, as we see by
rewriting the products in terms of sums of hyperbolic cosines of arguments $(\tau'-\tau)$ and $(\tau'+\tau+\beta)$.  The dependence on the latter cancels when we add (\ref{FACFI}) to (\ref{FACETA}). The sum may be written as \cite{hansen}:
\begin{equation}
\label{LINESUM}
\widetilde{G}_{\rm th}(\tau'-\tau;\bk)\equiv \frac{\cosh\{w_{\bk}[\beta/2-|\tau'-\tau|]\}}{2w_{\bk}\sinh(\beta w_{\bk}/2)}=\frac{1}{\beta}\sum_{n= -\infty}^{\infty}\frac{\cos[2\pi n T(\tau'-\tau)]}{(2\pi n T)^2+w_{\bk}^2}\;.
\end{equation}
It is worth noting that the first form of $\widetilde{G}_{\rm th}$ can be written in a familiar form which Pisarski\cite{pisa} has advocated for its practicality in calculations:
\begin{equation}
\label{PISARSKI}
\widetilde{G}_{\rm th}(\tau'-\tau;\bk)=\frac{1}{2w_{\bk}}[\frac{e^{-w_{\bk}|\tau'-\tau |}}{1-e^{-\beta w_{\bk}}}-\frac{e^{w_{\bk}|\tau'-\tau |}}{1-e^{\beta w_{\bk}}}].
\end{equation}
Clearly, adding the $\eta$ and $\phi$ lines reconstructs the thermal (finite $T$) propagator of the $d+1$-dimensional theory, which is $\tau$-translation invariant. Every graph in the expansion in vacuum bubbles of the original theory with the propagator (\ref{LINESUM}) may have each of its lines replaced by the sum of a $\eta$-line plus a $\phi$-line. This will generate all possible combinations of $\eta$ and $\phi$-lines which appear the (cutoff) reduced theory. This proves that integrating the reduced theory over $\phi$ will reproduce the results of perturbation theory to all orders, and that the only subtractions that are required are those of the $T=0$ theory in $d+1$ dimensions.

The final step to the DREA is very easy:  Instead of constructing the unadorned partition function by functional integration, as in equation (\ref{Z1}), we construct the partition function in the presence of a current $J(\bx )$:
\label{ZJ}
\begin{equation}
Z[\beta;J]\equiv \exp (-W[\beta;J]) =\int {\cal D} \phi \exp [-S_D+\int d^dx J(\bx )\phi (\bx )].
\end{equation}
Now all the integrations over both $\eta$ and $\phi$ have been done, so the same cancellations as in the partition function itself still occur.  One then only needs to Legendre-transform from $W[\beta;J]$ to the effective action $\Gamma [\beta;\langle\phi\rangle]$; this, the DREA, is free of spurious ultraviolet divergences.

Before applying these considerations to a specific field theory, we discuss how the general principles apply to gauge theories.

\section{Density matrix for gauge theories}
\label{gauge}

Begin by introducing notation.  We use the standard anti-Hermitean matrix form of the gauge potentials, with the coupling constant $g$ absorbed in the potential:
\begin{equation}
\label{notation}
gA_{\mu}(x)=\frac{\lambda_a}{2i}A_{\mu}^a;\;Tr\lambda_a\lambda_b=2\delta_{ab}.
\end{equation} 
Here the $\lambda_a/2$ are the group generators in conventional normalization.
Gauge transformations act on $A_{\mu}$ via the operation of unitary matrices $U$:
\begin{equation}
\label{ginv}
A'_{\mu}=UA_{\mu}U^{-1}+U\partial_{\mu}U^{-1}.
\end{equation}
It will sometimes be convenient to denote the time component of the $d+1$-dimensional potential $A_0$ as an adjoint scalar (in $d$ dimensions) field $\Phi$.  Greek indices refer to the $d+1$-dimensional theory and Latin indices to the dimensionally-reduced theory.
The potential $A_{\mu}$ depends on both $\tau$ and $\bx$.  There is a corresponding potential ${\cal A }_{\mu}$ depending only on $\bx$ which is the argument of the density matrix or of the DREA; this can also be written as $({\cal A }_i,\Phi)$.

The extension of our formalism to gauge theories requires some attention to questions of gauge invariance and gauge fixing.   The partition function can be written as the trace of $\exp (-\beta H)$ where $H$ is the microscopic time-independent Hamiltonian of the gauge theory, and the set of basis functions for the trace can be taken as the eigenfunctions of the functional Schr\"odinger equation.
Since $H$ commutes with the generator of time-independent gauge transformations, this basis may be chosen to be invariant under such gauge transformations.\footnote{At least for so-called small gauge transformations, without topological properties; we will not discuss what happens with large gauge transformations in this paper.} Therefore all matrix elements (of gauge-invariant operator products) found by tracing over the density matrix are gauge-invariant.   But this does not necessarily mean that it is convenient to calculate or present the density matrix as completely gauge-invariant; it proves more convenient to present it in a form where there is a gauge-fixing term.   This $d$-dimensional gauge-fixing term is inherited from a $d+1$-dimensional gauge-fixing term used to facilitate the gauge-theoretic analogs of the manipulations of the previous sections.  It turns out that the Feynman gauge is the simplest to use, and we will do so below.

The next question to discuss is that of periodicity of the gauge potential; in principle, periodicity need only be maintained up to a gauge transformation.  Because of the underlying gauge invariance of the basis used to calculate the trace in the partition function, one can (see, {\it e.g.}, \cite{gpy} for thermal gauge theory and \cite{cornwall} for the gauge-theory functional Schr\"odinger equation) introduce a projector which is an integral over all appropriate $d$-dimensional gauge transformations into the functional integrals.  Since small gauge transformations are generated by 
\begin{equation}
\label{omega}
\Omega\{\Lambda(\bx)\}=\exp [i\int d^dx Tr (\Lambda D_iE_i)]
\end{equation}
where $E_i$ is the electric field (and $i$ is a $d$-dimensional index), one sees the well-known fact that gauge invariance is the same as imposing Gauss' law on the physical states occurring in the partition-function trace.  As Ref. \cite{gpy} shows, inserting the projector
\begin{equation}
\label{proj}
P\equiv \int [{\cal D}\Lambda ]\Omega\{\Lambda(\bx)\}
\end{equation}
into the trace defining the partition function shows that the gauge partition function can be represented as a functional integral over all $d+1$ components of $A_{\mu}$ with strictly periodic boundary conditions, just as for the scalar case:
\begin{equation}
\label{pbc}
A_{\mu}(\beta ,\bx)=A_{\mu}(0,\bx ).
\end{equation}
After inserting the projector $P$ into the partition function, this quantity has a factor of the volume of the gauge transformations integrated over.  The usual Faddeev-Popov gauge-fixing procedure must be applied to isolate this factor, subject to one proviso.  These gauge transformations must obey a periodicity condition (which is simply periodicity of gauge-fixing ghosts):
\begin{equation}
\label{gf}
U(\beta ,\bx )=U_CU(0,\bx )
\end{equation}
where $U_C$ is an element of the center of the gauge group.  Non-trivial elements of the center are important for discussing such non-perturbative phenomena as center vortices and confinement, but this will not be taken up here.

Now we can show that using Feynman gauge allows for a trivial generalization of the construction of the density matrix for scalars.
In this gauge, the density-matrix propagator $G_\eta(\tau,\bx;\tau',\bx')_{\mu\nu}$ is very simply related to the scalar propagator of equation (\ref{GETA}):
\begin{equation}
\label{gaugeprop}
G_\eta(\tau,\bx;\tau',\bx')_{\mu\nu}=\delta_{\mu\nu}G_\eta(\tau,\bx;\tau',\bx').
\end{equation}

Similarly, the gauge potential $\widehat{A}_{\mu}(\beta ,\bx )=(A_i,\Phi)$ which enters into the construction of the densitiy matrix analogously to the field
$\fith (\beta ,\bx )$ of equation (\ref{PHITILDE}), is just:
\begin{equation}
\label{potential}
\widehat{ A}_{\mu}=\int \frac{d^d \bk}{(2\pi)^d}\, \frac{\cosh[w_{\bk}(\tau - \beta/2)]}{\cosh(\beta w_{\bk}/2)}\,
{\widetilde{\cal A }}_{\mu}(\bk )  
\end{equation}
where ${\widetilde{\cal A} }_{\mu}(\bk )$ is the Fourier transform of the field ${\cal A }_{\mu}(\bx )$ which is the argument of the density matrix.  Repeating the arguments of Section \ref{uvbehavior} then shows that the thermal propagator for the gauge theory is again related to that of the scalar theory by a factor of $\delta_{\mu\nu}$.

The free term of the density matrix, constructed analogously to the scalar field density matrix (see equation (\ref{SFI})), is found to contain the Feynman gauge-fixing term in $d$ dimensions, as expected.  As mentioned above, it is also necessary to include periodic ghost contributions whose construction is entirely parallel to that of the scalar fields discussed above.  Needless to say, the whole procedure can be carried out in any gauge, but certainly the simplest presentation of the technique is in the Feynman gauge.  

We will postpone applications of this formalism in gauge theories to future work, and now turn to a simple application in a scalar field theory.


\section{Applications to scalar theories}
\label{applications}

The results of Section \ref{densitymatrix} provide us with a constructive method for deriving a DREA that can be interpreted as equivalent to the Landau ``coarse-grained free energy" of the original microscopic theory. The $T$ dependence of this free energy can be derived order by order. As long as we keep a cutoff (or use some other regularization method), we can proceed to compute correlations in the same manner as in any ordinary statistical mechanics problem. The renormalization required to obtain physical quantities can performed at the very last step, and will reflect the ultraviolet
behavior of the original theory. On the other hand, the physical quantities themselves will be those of a $d$-dimensional theory whose statistical weights are dictated by the exponential of our reduced action. 

For practical applications, it is convenient to use the reduced theory not only because it reduces the number of dimensions, but also because it allows us to make contact with the vast literature on statistical mechanics. In this Section, we will illustrate how this comes about.
  

\subsection{The DREA for $\varphi^4$ theory}
\label{efftheory}

In order to compute correlations, we need not literally go through the steps described in Section 2 of adding sources $J$ and integrating over both the $\eta$ and $\phi$ fields, followed by Legendre transformation, to actually find the DREA.  A minor shortcut consists of obtaining the Feynman rules which emerge as we match the $\phi$ integrations with the $\eta$ integrations, resulting in a cancellation of ultraviolet divergences. For example, if we choose to compute $\langle \fiti(\bp_1)\fiti(\bp_2) \rangle$, $\bp_1=-\bp_2=\bp$, the combination of the $\eta$ and $\phi$ propagators which appear as internal lines in the various diagrams as a result of the integrations over those fields leads to a cancellation of divergences, and to the appearance of the usual thermal propagator  $\widetilde{G}_{\rm th}$. As we will argue below, the resulting graphically-defined  theory yields, for $\langle \fiti(\bp_1)\fiti(\bp_2) \rangle$, the sum of all graphs that would appear in a $\phi^4$ theory in $d$-dimensions, with each internal line of momentum $\bk$ corresponding to $\widetilde{G}_{\rm th}(\tau_2-\tau_1,\bk)$ (the  thermal propagator), and with each external line given by the $\widetilde{G}_{\phi}(\bp_j)$ propagator ($j=1,2$) of equation (\ref{GFI}) multiplied by the factor $\cosh[w_{\bp_j}(\tau_j - \beta/2)]/\cosh(\beta w_{\bp_j}/2)$, all this integrated over $\tau_1$ and $\tau_2$. This is the field theory defined by the DREA.

  In order to show that the perturbative rules for the DREA are indeed the ones mentioned in the previous paragraph, we will consider the $n$-point correlation
\begin{equation}
\label{CORREL}
\langle \fiti(\bp_1)\cdots\fiti(\bp_n) \rangle \equiv Z^{-1}(\beta)\int [{\cal D}\phi] 
\,\fiti(\bp_1)\cdots\fiti(\bp_n)\,\rho[\beta;\phi,\phi]\;,
\end{equation}
where $\rho[\beta;\phi,\phi]$ is given by (\ref{RHO}). If we expand the interacting part, and make use of (\ref{ETAETA}) and (\ref{SUMP}) in the integral over $\eta$, these $\eta$-$\eta$ contractions will generate graphs with internal lines which correspond to $\widetilde{G}_{\eta}$, leaving the $\fith$ and $\phi$ fields uncontracted. The final integral over $\phi$ is to be performed with the Gaussian weight $e^{S_F[\phi]}$, given by (\ref{SFI}). This last integral will introduce $\fith$-$\fith$, and $\fith$-$\phi$ contractions. The former yield factors like those in (\ref{FACFI}), whereas the latter produce factors such as
\begin{equation}
\label{FACTORS}
\widetilde{G}_{\phi}(\bp_j)\frac{\cosh[w_{\bp_j}(\tau_j - \beta/2)]}{\cosh(\beta w_{\bp_j}/2)}\;,
\end{equation}
for $j=1,\ldots,n$. Now, the sum of (\ref{FACFI}) and (\ref{FACETA}) gives (\ref{LINESUM}); to every $\eta$-$\eta$ contraction there is a corresponding $\fith$-$\fith$ contraction; and their weights are identical, as they come from the $\varphi^4$ graphs, with $\varphi=\fith+\eta$. All this shows that we shall have the graphs of $\phi^4(\bx)$ theory, with $\widetilde{G}_{\rm th}$ internal propagators, and external lines given by (\ref{FACTORS}), with the $\tau$ dependences of $\widetilde{G}_{\rm th}$ and (\ref{FACTORS}) integrated over.

In the sequel, we will give an explicit example of how this works by examining the case of broken $\varphi^4$ theory and performing a first order calculation.      


\subsection{The phase transition in $d=2$}
\label{phase}

We will illustrate at one-loop level how the concept of dimensional reduction via the density matrix may be used to study the phase structure of $\varphi^4$ theory, as an example of a physical application. We will not be concerned with the fact that a one-loop calculation may not be an accurate description as we approach a phase transition, but will use this as an illustration of the techniques to compute physical masses in our scheme. Accurate or not, we will derive a criterion for a phase transition by demanding that the physical mass, the inverse of a correlation length, vanish at the transition point.

The first nontrivial example we can treat is the case $d=2$. The case $d=0$ is just quantum mechanics, its dimensional reduction leading to ordinary integrals, whereas the $d=1$ case can be reduced to an effective quantum-mechanical problem, where no transition takes place. For $d=2$ the theory is super-renormalizable, and only requires a mass renormalization. It has been studied in approximations involving both lattice computations \cite{amaral,spanish} and analytic methods \cite{wipf,ej,lichen,larry}. Our discussion will make contact with both.

We start from an interacting Lagrangean defined by:
\begin{equation}
\label{BROFI4}
{\cal L}_I[\varphi]=(\lambda/4!)(\varphi^2-\varphi_0^2)^2 - (1/2)m^2\varphi^2\;.
\end{equation}
With real $m$ the last term guarantees that the cutoff theory is in the broken symmetry phase. In order to derive the physical mass, we will compute the tadpole contribution to $\langle\phi(\bp)\phi(-\bp)\rangle$, and minimally subtract the equivalent term of the $T=0$ theory (ie, the tadpole in $d+1=3$). The self-energy, to first order, is given by:
\begin{equation}
\label{SELF}
\Sigma_\Lambda(\beta;\bp)=F(\beta;\bp)\left[(m^2+\lambda\varphi_0^2)-\frac{\lambda}{2}\int \frac{d^d \bk}{(2\pi)^2} \widetilde{G}_{\rm th}(0;{\bk})\right]\;,
\end{equation}
\begin{equation}
\label{FP}
F(\beta;\bp)\equiv \frac{\beta}{2\cosh^2(\beta w_{\bp}/2)}[\, 1+ \frac{\sinh(\beta w_{\bp})}{\beta w_{\bp}}]\;,
\end{equation}
where the last term of (\ref{SELF}) is the result of adding the tadpole with internal $\eta$ lines (which appears in the Appendix) to the one coming from contracting the first term in (\ref{RHO1}) of the Appendix with the $\phi$ propagator. In fact, the sum cancels the second term in (\ref{ID}) of the Appendix, restores $\tau$-translation invariance, and eliminates the $d=2$ divergences explicitly shown in (\ref{I'}, \ref{I''}, \ref{I'''}).

We now choose $\Sigma_\Lambda(\beta;\bp)=0$ for $\beta\to\infty $ so that $m$ is the physical mass parameter of the zero-temperature theory. Since $F\to (1/w_{\bp})$, we may infer the $\Lambda$-dependence to be attributed to $\lambda\varphi^2_0$ in order to eliminate the $T=0$ divergence. Using the relation:
\begin{equation}
\label{SPLIT}
\frac{1}{2w_{\bk}\tanh(w_{\bk}\beta/2)}=\frac{1}{2w_{\bk}} + \frac{1}{w_{\bk}(e^{w_{\bk}\beta}-1)}\;,
\end{equation}
which splits the $T=0$ and $T\neq 0$ parts of the last term of (\ref{SELF}), and $w_{\bk}dw_{\bk}=kdk$, we find that the divergence goes like $\Lambda$, whereas the finite remainder $\Sigma_R$ contributes:
\begin{equation}
\label{SELFR}
\Sigma_R=\frac{\lambda T}{4\pi}\ln(1-e^{-m/T})\;.
\end{equation}
The $T$-dependent correlation length $\xi=(m^2+\Sigma_R)^{-1/2}$ becomes infinite whenever $T$ obeys:
\begin{equation}
\label{TRANS}
m^2=-\frac{\lambda T}{4\pi}\ln(1-e^{-m/T})\;,
\end{equation}
which, for $T\gg m$, becomes:
\begin{equation}
\label{TRANS2}
m^2=\frac{\lambda T}{4\pi}\ln(\frac{T}{m})\;.
\end{equation}
There is always a critical temperature $T_c$ satisfying this equation for any positive value of $\lambda$, with $T_c\gg m$ if $\lambda \ll m$.
We emphasize that (\ref{SELF})  is exactly the same physical mass as occurs in the DREA as calculated from $Z[\beta;J]$, as claimed in \ref{efftheory}. 

The discussion above has centered on rules for finding the DREA, or equivalently the matrix elements defined by the DREA.  Once the DREA has been found, the whole theory is known and no further work is necessary.  However,
it is easier, in principle at least, to calculate the density matrix (or the corresponding  action $S_D$ of equation (\ref{RHO})) by only integrating over the $\eta$ fields, postponing the $\phi$ integrations to a later step, such as numerical integration of the theory defined by $S_D$ on a lattice.  In such a case there will remain spurious ultraviolet divergences, which must go away when matrix elements are computed via the $\phi$ functional integrals.
This can be compared to a theory defined on a two-dimensional lattice. Its parameters depend on the physical $T$ and on the cutoff (ie, the lattice spacing). We may then obtain the temperature where symmetry restoration occurs as a function of the cutoff. Removing the cutoff, so as to recover the continuum limit, will yield the physical transition $T$ of the continuum theory. The calculation we have shown corresponds to a one-loop realization of this process; according to Refs. \cite{spanish,ej}, this is not terribly bad numerically.


\section{Conclusions}
\label{conclusions}

We first introduced the notion of using the (negative logarithm of) the density matrix as defining an action $S_D$ which would be a useful replacement for the usual DRFT, since it would not only be a theory in a reduced dimension ($d+1\rightarrow d$) but it would also be defined at all temperatures and might furnish a workable definition of the Landau-Ginzburg free energy, complete with information on phase transition temperatures coming from integrating over the Euclidean time dynamics.  It then appeared that the action $S_D$ could have spurious ultraviolet divergences not allowed in the partition function and its matrix elements, since these spurious divergences were absent in the $T=0$ theory.
We showed that one could define a dimensionally-reduced effective action, the DREA, which was free of these spurious divergences as a result of cancellations between functional integrals over fields defining the Euclidean time evolution and functional integrals over the argument fields of $S_D$. 
The DREA is more difficult to calculate than the density matrix, since as an effective action it has infinitely-many terms involving all the Green's functions of the full theory..  We have discussed here the graphical construction of the DREA correlations at the one-loop level. It is clear that dressed-loop techniques \cite{cornwall,cjt} apply for the DREA just as they do for any effective action; this allows for non-perturbative phenomena such as phase transitions to be investigated with the DREA.  For example, corrections to the thermal propagator $\widetilde{G}_{\rm th}(\tau_2-\tau_1,\bk)$ can be formally summed so that the DREA is expressed entirely in terms of the dressed thermal propagator.         

We saw that one important difference between the DRFT and the DREA theory is that the DRFT may have ultraviolet divergences which are not inherent in the finite-$T$ version of the original $d+1$ theory.  These appear as finite and calculable terms, generally involving $\ln T$, in the DREA theory and become the DRFT divergences at infinite temperature.  We have studied an example of such divergences in the $d=2+1$ $\varphi^4$ model, studied by Einhorn and Jones \cite{ej} using the DRFT.  A one-loop mass divergence in the DRFT is replaced by a finite $\ln T$ term in the DREA theory.  A very similar phenomenon takes place in $d=2+1$ non-Abelian gauge theory, whose DRFT is $d=2$ gauge theory coupled to an adjoint scalar (the original time component of the gauge potential).  This scalar in the DRFT has a logarithmic mass divergence which is actually a $\ln T$ term at finite temperature, as calculated long ago by D'Hoker \cite{dh}.  The logarithmic divergence plays a very interesting role in understanding the transition from $d=2+1$ gauge theory at zero $T$ to the infinite-temperature limit, since the $d=2+1$ theory has zero string tension for adjoint and similar Wilson loops but the corresponding DRFT has non-vanishing string tension for all representations of the Wilson loop.  This question is under study by two of us \cite{ca}.  There will be other important uses of the DREA in investigations of $d=3+1$ gauge theories as well (see, {\it e.g.}, \cite{drdisc}).

We have emphasized that the DREA is capable of answering, in principle at least, questions about condensed-matter phase transitions which are not within the purview of the usual renormalization-group approach to second-order phase transitions, such as the value of the phase transition temperature.  Evidently, one aspect of the extension of the DREA to a dressed-loop expansion will continue to be the use of the renormalization group, so that one hopes to get both critical exponents and phase-transition temperatures from the DREA.  Ultimately one hopes that the DREA can be truncated to relevant and marginal operators (for purposes of critical exponents) plus perhaps one or two more terms in the DREA expansion.  

Finally, we have noted but not explored at all the fact that the extension of the techniques used here for the density matrix can be straightforwardly extended to objects like the Wigner distribution function, which amounts to calculating off-diagonal matrix elements. 
This will be further explored in subsequent papers.


\begin{center}
{\bf Acknowledgements}
\end{center}

The authors acknowledge support from CNPq (CAAC and AJS), FAPERJ and FUJB/UFRJ(CAAC), and FAPESP(AJS). CAAC and AJS would like to thank UCLA for its hospitality. JMC thanks the Aspen Center for Physics, where part of this work was done.

\appendix
\section{Appendix}

Let us compute the contribution to the reduced action in first order for $d=2$, in order to illustrate how divergences appear.
From expression (\ref{RHOA}), we have: 
\begin{equation}
\label{RHO1}
\rho_1[\beta;\phi,\phi]=-\frac{\lambda}{4!}\rho_0\,\left[\int_1 \widehat\varphi^4(1)+
6\int_1 G_\eta(1;1)\widehat\varphi^2(1)+\int_1 \langle\eta^4(1)\rangle \right]\;.
\end{equation}
The last term in (\ref{RHO1}) is independent of $\fith$. The term in $\fith^2$ corresponds to a tadpole graph which we will call $T_\eta$. Then, using (\ref{PHITILDE}), (\ref{PHIK}) and (\ref{GETA}) we arrive at:
 %
\begin{equation}
\label{ETATAD}
T_\eta=\int \frac{d^d \bp}{(2\pi)^d}\, \frac{\phi(\bp)\phi(-\bp)}{\cosh^2(\beta w_{\bk}/2)}\, \int \frac{d^d \bk}{(2\pi)^d}\, I_\eta(\bp,\bk)\;,
\end{equation}
\begin{equation}
\label{IETA}
 I_\eta(\bp,\bk)=\int_0^\beta d\tau \cosh^2[w_{\bp}(\tau-\beta/2)] \frac{\sinh(w_{\bk}\tau)\sinh[w_{\bk}
(\beta-\tau)]}{w_{\bk}\sinh(\beta w_{\bk})} \;.
\end{equation}
%
The integral over $\tau$ is straightforward. It is convenient to use the identity:
\begin{equation}
\label{ID}
\sinh(w_{\bk}\tau)\sinh[w_{\bk}(\beta-\tau)]=\cosh^2(\beta w_{\bk}/2)-\cosh^2[w_{\bk}(\tau-\beta/2)]\;, 
\end{equation}
and to split $I_\eta$ into a sum of three contributions, $I'_\eta$, $I''_\eta$ and $I'''_\eta$. Explicitly:
%
\begin{equation}
\label{I'}
I'_\eta=\frac{\beta[1+\sinh(\beta w_{\bp})/(\beta w_{\bp})]}{4w_{\bk}\tanh(\beta w_{\bk}/2)}\;,
\end{equation}
\begin{equation}
\label{I''}
I''_\eta=-\frac{1}{4w_{\bk}^2}-\frac{\sinh[\beta (w_{\bp}+w_{\bk})]}{8w_{\bk}(w_{\bp}+w_{\bk})\sinh(\beta w_{\bk})}- \frac{\sinh[\beta(w_{\bp}-w_{\bk})]}{8w_{\bk}(w_{\bp}-w_{\bk})\sinh(\beta w_{\bk})}\;,
\end{equation}
\begin{equation}
\label{I'''}
I'''_\eta = -\frac{\beta[1+\sinh(\beta w_{\bp})/(\beta w_{\bp})]}{4w_{\bk}
\sinh(\beta w_{\bk})}\;.
\end{equation}
%
In $d=2$, we use $w_{\bk}dw_{\bk}=kdk$. The integrals go from $|m|$ to $(\Lambda^2+m^2)^{1/2}$. The contribution from $I'_\eta$ goes like $\Lambda$, the one from $I''_\eta$ goes like $\ln\Lambda$, whereas the one from $I'''_\eta$ is finite as $\Lambda\to\infty$. In fact, the two contributions that diverge for large $\Lambda$ correspond to the tadpoles in $d=3$ and $d=2$, respectively, if we use ${\bk}^2+m^2$ as the inverse propagator. Although one might feel tempted to use these two types of subtractions to arrive at finite results, this is not
what has to be done as shown in Section \ref{uvbehavior}.


\end{document}